\documentclass[10pt,conference]{IEEEtran}

\usepackage{amsmath, graphicx, amssymb, amsfonts, subfigure, arydshln, setspace, color, epsfig,graphicx,epstopdf}

\newcommand{\erase}[1]{{\color{blue}TED Erase:}}

\newtheorem{dfn}{Definition}

\newtheorem{theorem}{Theorem}

\newtheorem{exam}{Example}

\renewcommand{\geq}{\geqslant}

\newcommand{\highsup}[1]{\raisebox{0.35ex}{\kern 1pt $\scriptstyle {#1}
$}}

\newcommand{\ec}{\end{center}}

\outer\def\proclaim #1. #2\par{\medbreak
 \noindent{\bf#1.\enspace}{\sl#2\par}%
 \ifdim\lastskip<\medskipamount \removelastskip\penalty55\medskip\fi}

\makeatletter

\gdef\@punct{.\ \ }  
\def\@sect#1#2#3#4#5#6[#7]#8{%
  \ifnum #2>\c@secnumdepth
     \def\@svsec{}
  \else
     \refstepcounter{#1}\edef\@svsec{%
     \ifnum #2>0{{\csname the#1\endcsname}}.\fi%
    \hskip .5em}
  \fi
  \@tempskipa #5\relax
  \ifdim \@tempskipa>\z@
     \begingroup #6\relax
       \@hangfrom{\hskip #3\relax\@svsec}{\interlinepenalty \@M #8\par}
     \endgroup
     \csname #1mark\endcsname{#7}
     \addcontentsline{toc}{#1}{\ifnum #2>\c@secnumdepth\else
          \protect\numberline{\csname the#1\endcsname}\fi#7}
  \else
     \def\@svsechd{#6\hskip #3\@svsec #8\@punct\csname
#1mark\endcsname{#7}
     \addcontentsline{toc}{#1}{\ifnum #2>\c@secnumdepth \else
          \protect\numberline{\csname the#1\endcsname}\fi#7}}
  \fi
  \@xsect{#5}}

\def\@ssect#1#2#3#4#5{\@tempskipa #3\relax
  \ifdim \@tempskipa>\z@
     \begingroup #4\@hangfrom{\hskip #1}{\interlinepenalty \@M
#5\par}\endgroup
  \else \def\@svsechd{#4\hskip #1\relax #5\@punct}\fi
  \@xsect{#3}}

\makeatother

\begin{document}

\allowdisplaybreaks

\title{Network Equivalence in the Presence\\ of an Eavesdropper}

\author{
\authorblockA{$\text{Theodoros K. Dikaliotis}^1$, $\text{Hongyi Yao}^2$ $\text{Tracey Ho}^1$, $\text{Michelle Effros}^1$, $\text{Joerg Kliewer}^3$\\
$\hspace{1mm}^{1,2}$California Institute of
Technology $\hspace{1mm}^3$New Mexico State University\\
{$^1$\{tdikal, tho, effros\}@caltech.edu $\hspace{1mm}^2$yaohongyi03@gmail.com
$\hspace{1mm}^3$jkliewer@nmsu.edu}}}

\maketitle

\begin{abstract}
We  consider networks of noisy degraded wiretap channels in the presence of an eavesdropper. For  the case where the eavesdropper can wiretap at most one channel at a time, we show that the secrecy capacity region, for a broad class of channels and any given network topology and communication demands, is equivalent to that of a corresponding network where each noisy wiretap channel is replaced by a noiseless wiretap channel. Thus in this case there is a separation between wiretap channel coding on each channel and secure network coding on the resulting noiseless network.  We show with an example that such separation does not hold when the eavesdropper can access multiple channels at the same time, for which case we provide upper and lower bounding noiseless networks.
\end{abstract}

\section{Introduction}
\label{sec:introduction}

Information theoretically secure (secret) communication in the presence of an eavesdropper has been studied under various models. One body of literature studies the wiretap channel, introduced by Wyner~\cite{wyner75}, where the intended receiver and the eavesdropper observe outputs of a physical layer channel. 
Another body of literature investigates the secure capacity of networks of noise-free links. Under this model, introduced by Cai and Yeung in~\cite{cai02secure}, an eavesdropper perfectly observes all information traversing a restricted but unknown subset of links. 
The first paper on the secure capacity of a network of noisy channels is~\cite{mills08}, which finds upper and lower bounds on the unicast capacity of a network of independent broadcast erasure channels when the output observed by the eavesdropper equals that of the intended receiver on all wiretapped channels.

Our work considers the problem of secure communication over a network of independent wiretap channels which are physically degraded and ``simultaneously maximizable''  (see Definition~\ref{dfn:simultaneously_mutual_information_maximizable} in Section~\ref{sec:Network_model}), and  
broadens  consideration to general capacity regions specifying vectors of simultaneously achievable rates. We require asymptotically negligible decoding error probability and information leakage to the eavesdropper, as defined formally in Section~\ref{sec:Network_model}. 
In the case where the eavesdropper has access to only one link, the identity of which is unknown to the code designer, we show that the secrecy capacity region is identical to that of a corresponding noiseless network, for any network topology and connection types. Thus in this case capacity can be achieved by separate design of wiretap channel codes converting each channel to a pair of public and confidential noiseless links, and a secure network code on the resulting noiseless network.  We show with an example that such separation does not hold when the eavesdropper can access multiple channels at the same time, for which case we provide upper and lower bounding noiseless networks.
Our results bring together and generalize the wiretap channel and secure network coding literature, allowing application of existing results on secure network coding capacity  to  characterize or bound the secure capacity of networks of such wiretap channels. 
Our work builds on and generalizes the techniques developed by Koetter, Effros, and Medard in~\cite{network_equiv_partI, network_equiv_partII}, which show similar capacity bounds in the absence of secrecy constraints. We provide below outlines of all proofs, details of which are   given in the full version of this paper~\cite{eav_equvpreprint}.

\section{Model and Preliminaries}
\label{sec:Network_model}
 Consider a network $\mathcal{G}=(\mathcal{V},\mathcal{E})$, where
$\mathcal{V}$ is the set of nodes and $\mathcal{E}\subseteq\mathcal{V}\times\mathcal{V}\times\mathbb{N}$ is a set of directed edges between pairs of nodes in the network. Edge $(i,j,k)$ represents the $k^{\text{th}}$ wiretap channel through which node $i$ communicates to node $j$ and through which an eavesdropper may or may not be listening. The total number of nodes in the network is $m$. The channel inputs and outputs for node $i$ at time $t$ are given by
\begin{align*}
X_t^{(i)}=\left(X_t^{(e)}:e\in\mathcal{E}_\text{out}(i)\right)\quad\text{and}\quad
Y_t^{(i)}=\left(Y_t^{(e)}:\mathcal{E}_\text{in}(i)\right)
\end{align*}
where $X_t^{(e)}$ and $Y_t^{(e)}$ denote the input to and the output from edge $e$ respectively,  and $\mathcal{X}^{(e)}$ and $\mathcal{Y}^{(e)}$ denote their alphabets, which may be discrete or continuous. 
We define  
\begin{align*}
\mathcal{E}_\text{in}(i) &= \left\{(u, v, w)\in\mathcal{E}: v = i\right\}\\
\mathcal{E}_\text{out}(i) &= \left\{(u, v, w)\in\mathcal{E}: u = i\right\}
\end{align*}
%
%
\begin{align*}
\mathcal{X}^{(i)}=\prod_{e\in\mathcal{E}_\text{out}(i)}\mathcal{X}^{(e)}\quad \text{and}\quad\mathcal{Y}^{(i)}=\prod_{e\in\mathcal{E}_\text{in}(i)}\mathcal{Y}^{(e)}.
\end{align*}

Let $\mathcal{P}(\mathcal{E})$ denote the power set of the set of all edges. In a secure communication problem,  an adversarial set $A\subseteq\mathcal{P}(\mathcal{E})$ is specified. Each set $E\in A$ describes a subset of channels over which an eavesdropper may be listening. The  code is designed to be secure against eavesdropping on the set of channels $E$ for every $E\in A$. When the eavesdropper listens to edge $e = (i,j,k)$, the eavesdropper receives, at each time $t$, a degraded version $Z^{(e)}_t$ of the channel output $Y^{(e)}_t$ observed by the intended recipient, which is the output node $j$ of edge $e=(i,j,k)$. If the eavesdropper has eavesdropping set $E\in A$, then at time $t$ it receives the set of random variables $\left(Z^{(e)}_t:e\in E\right)$, which we compactly write as $Z_t^{(E)}$. The vector $\left(Z_1^{(E)},\ldots,Z_n^{(E)}\right)$ of observations from all edges $e\in E$ over time steps $t\in\{1,\ldots,n\}$ is denoted by $\left(Z^{(E)}\right)^n$. Similarly we define $\left(X^{(E)}\right)^n=\big(X^{(E)}_1,\ldots,X^{(E)}_n\big)$ and $\left(Y^{(E)}\right)^n=\big(Y^{(E)}_1,\ldots,Y^{(E)}_n\big)$ where $X^{(E)}_t = \left(X^{(e)}_t:e\in E\right)$ and $Y^{(E)}_t = \left(Y^{(e)}_t:e\in E\right)$.

For each $e\in\mathcal{E}$, channel $e$ is a memoryless, time-invariant, physically degraded wiretap channel described by a conditional distribution
\begin{align*}
p(y^{(e)},z^{(e)}|x^{(e)})=p(y^{(e)}|x^{(e)})\cdot p(z^{(e)}|y^{(e)}).
\end{align*}
All wiretap channels are independent by assumption, giving
\begin{align*}
&p\big(y^{(\mathcal{E})},z^{(\mathcal{E})}|x^{(\mathcal{E})}\big)=\prod_{e\in\mathcal{E}}p\big(y^{(e)},z^{(e)}|x^{(e)})\\=&\prod_{e\in\mathcal{E}}p\big(y^{(e)}|x^{(e)})p\big(z^{(e)}|y^{(e)}).
\end{align*}
We further restrict our attention to channels that are ``simultaneously maximizable,'' as defined below.
\begin{dfn}
Wiretap channel $e$ is called simultaneously maximizable if \begin{align*}\arg\left[\mathop{\max}_{p(x)}I(X^{(e)};Y^{(e)})\right] = \arg\left[\mathop{\max}_{p(x)}I(X^{(e)};Z^{(e)})\right]
\end{align*}
 and
\begin{align*}
&\displaystyle\mathop{\max}_{p(x^{(e)})}\left[I(X^{(e)};Y^{(e)})-I(X^{(e)};Z^{(e)})\right] \\=&\displaystyle\mathop{\max}_{p(x^{(e)})}I(X^{(e)};Y^{(e)})-\displaystyle\mathop{\max}_{p(x^{(e)})}I(X^{(e)};Z^{(e)}).
\end{align*}
\label{dfn:simultaneously_mutual_information_maximizable}
\end{dfn}
The about maximization 
is subject to any constraints on the channel input (\emph{e.g.}, an input power constraint for a Gaussian channel) associated with the communication network of interest. Examples of simultaneously maximizable wiretap channels include weakly symmetric channels and Gaussian channels~\cite{Gaussian_wiretape,Fadding_wiretape}.
Intuitively, restriction to simultaneously maximizable channels simplifies our analysis since the same input distribution maximizes both the total and secure capacity. 

A code of blocklength $n$ operates over $n$ time steps to reliably communicate 
message
\begin{align*}
W^{(i\rightarrow\mathcal{B})}\in\mathcal{W}^{(i\rightarrow \mathcal{B})}\displaystyle\mathop{=}^\text{def}\{1,\ldots,2^{nR^{(i\rightarrow \mathcal{B})}}\}
\end{align*}
from each source node $i\in\mathcal{V}$ to each nonempty set $\mathcal{B}\subseteq\mathcal{V}\backslash\{i\}$ of sink nodes in a manner that guarantees information theoretic security in the presence of any eavesdropper $E\in A$. This  constitutes a unicast connection if $|\mathcal{B}|=1$ and a multicast connection if $|\mathcal{B}|>1$. Constant $R^{(i\rightarrow\mathcal{B})}$ is called the transmission rate from source $i$ to sink set $\mathcal{B}$.  The vector of all rates $R^{(i\rightarrow\mathcal{B})}$ is denoted by $R=\left(R^{(i\rightarrow \mathcal{B})}:i\in\mathcal{V}, \mathcal{B}\in\mathcal{B}^{(i)}\right)$, where set $\mathcal{B}^{(i)}=\left\{\mathcal{B}:\mathcal{B}\subseteq\mathcal{V}\backslash\{i\}, \mathcal{B}\neq\emptyset\right\}$ is the set of non-empty receiver sets to which node $i$ may wish to transmit. Similarly, the vector of all messages is denoted by $W=\left(W^{(i\rightarrow\mathcal{B})}:i\in\mathcal{V}, \mathcal{B}\in\mathcal{B}^{(i)}\right)$.

Each node $i\in\mathcal{V}$ also has access to a random variable $T^{(i)}\in\mathcal{T}^{(i)}\displaystyle\mathop{=}^{\text{def}}\{1,\ldots,2^{nC^{(i)}}\}$ for use in randomized coding for secrecy, where
\begin{align} C^{(i)}=\displaystyle\sum_{e\in\mathcal{E}_{\text{out}}(i)}\max_{p(x^{(e)})}I\big(X^{(e)};Y^{(e)}\big)
\label{eq:outgoint_channel_capacity}
\end{align}
is the sum of the outgoing channel capacities from node $i$. 
Each $T^{(i)}$ is uniformly distributed on its alphabet and independent of all messages and channel noise.
\begin{dfn}
Let a network
\begin{align*}
&\mathcal{N}\displaystyle\mathop{=}^\text{def}(\prod_{e\in\mathcal{E}}\mathcal{X}^{(e)}, \prod_{e\in\mathcal{E}}\Big(p(y^{(e)}|x^{(e)}) p\left(z^{(e)}|y^{(e)}\right)\Big),\\& \prod_{e\in\mathcal{E}} \left(\mathcal{Y}^{(e)}\times\mathcal{Z}^{(e)}\right))
\end{align*}
be given corresponding to a graph $\mathcal{G}=(\mathcal{V}, \mathcal{E})$. A blocklength $n$ solution $\mathcal{S}(\mathcal{N})$ 
is defined as a set of encoding functions
\begin{align*}
X^{(i)}_t &:\left(\mathcal{Y}^{(i)}\right)^{t-1}\times\prod_{\mathcal{B}\in\mathcal{B}^{(i)}}\mathcal{W}^{(i\rightarrow \mathcal{B})}\times\mathcal{T}^{(i)}\longrightarrow \mathcal{X}^{(i)}
\end{align*}
mapping $\left(Y^{(i)}_1,\ldots, Y^{(i)}_{t-1}, \big(W^{(i\rightarrow\mathcal{B})}:\mathcal{B}\in\mathcal{B}^{(i)}\big),T^{(i)}\right)$ to $X^{(i)}_t$ for each $i\in\mathcal{V}$ and $t\in\{1,\ldots,n\}$, and a set of decoding functions
\begin{align*}
\breve{W}^{(j\rightarrow\mathcal{K}, i)} &:\left(\mathcal{Y}^{(i)}\right)^n\times\prod_{\mathcal{B}\in\mathcal{B}^{(i)}}\mathcal{W}^{(i\rightarrow \mathcal{B})}\times\mathcal{T}^{(i)}\longrightarrow \mathcal{W}^{(j\rightarrow\mathcal{K})}
\end{align*}
 mapping $\left(Y^{(i)}_1,\ldots, Y^{(i)}_n,\big(W^{(i\rightarrow\mathcal{B})}: \mathcal{B}\in\mathcal{B}^{(i)}\big),T^{(i)}\right)$ to $\breve{W}^{(j\rightarrow\mathcal{K}, i)}$ for each $j\in\mathcal{V}$, $\mathcal{K}\in\mathcal{B}^{(j)}$, and $i\in\mathcal{K}$. The solution $\mathcal{S}(\mathcal{N})$ 
 is called a $(\lambda,\varepsilon,A,R)$--solution, denoted $(\lambda,\varepsilon,A,R)$--$\mathcal{S}(\mathcal{N})$, if 
 $\Pr\left(\breve{W}^{(j\rightarrow\mathcal{K}, i)}\neq W^{(j\rightarrow\mathcal{K})}\right)<\lambda$ for every $j\in\mathcal{V}$, $\mathcal{K}\in\mathcal{B}^{(j)}$ and $i\in\mathcal{K}$, and $I\left(\left(Z^E\right)^n;W\right)<n\varepsilon$ for every $E\in A$.
\label{dfn:defining_a_network}
\end{dfn}

\begin{figure}
\centering
\includegraphics[width=0.45\columnwidth]{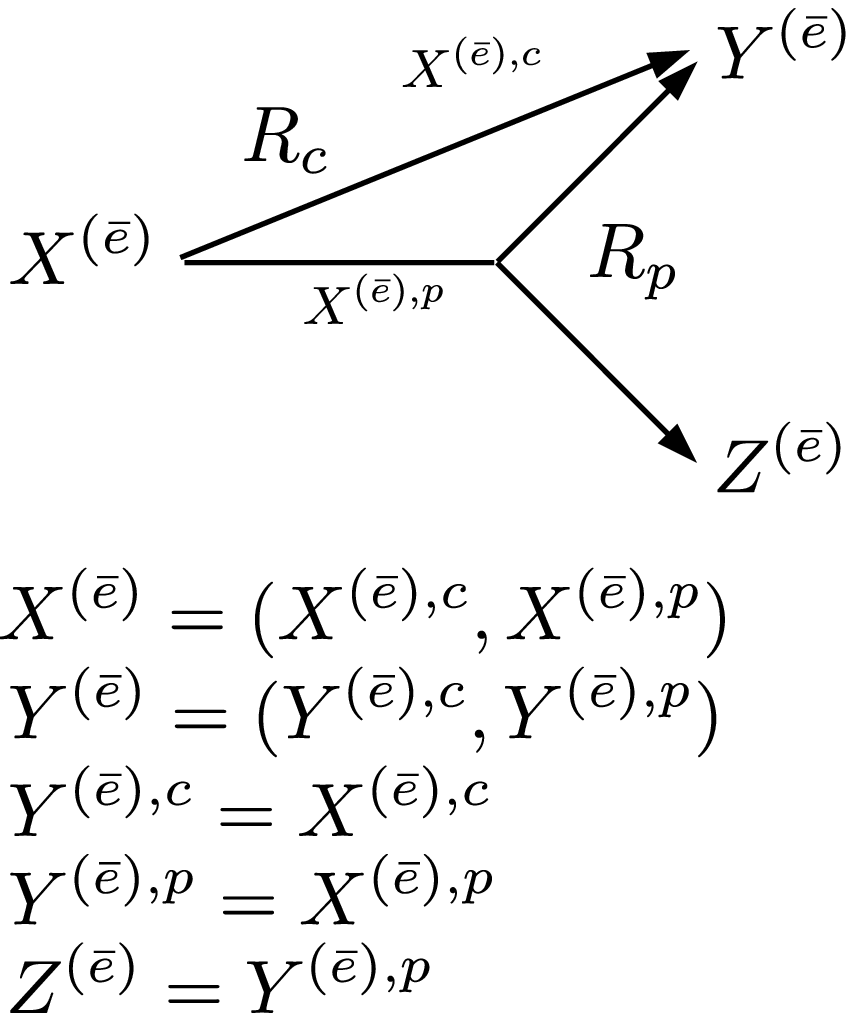}
\caption{A noiseless degraded broadcast channel with confidential rate $R_c$ and public rate $R_p$.}
\label{fig:rate_regions_noiseless}
\end{figure}

\begin{dfn}
The $A$--secure rate region $\mathcal{R}(\mathcal{N},A)\subseteq \mathbb{R}\hspace{0.4mm}^{m(2^{m-1}-1)}_+$ of a network $\mathcal{N}$ is the closure of all rate vectors $R$ such that for any $\lambda>0$ and $\varepsilon>0$, a solution $(\lambda,\varepsilon,A, R)$--$\mathcal{S}(\mathcal{N})$ exists.
\label{dfn:defining_a_rate_region}
\end{dfn}

Given a network $\mathcal{N}$ and a  channel $\bar{e}\in\mathcal{E}$, the model $\mathcal{N}_{\bar{e}}(R_c,R_p)$ 
replaces $\bar e$ with noiseless bit pipes as defined below and illustrated in Figure~\ref{fig:rate_regions_noiseless}.
\begin{dfn}
Given a network
\begin{align*}
&\mathcal{N}\displaystyle\mathop{=}^\text{def}(\prod_{e\in\mathcal{E}}\mathcal{X}^{(e)},\prod_{e\in\mathcal{E}}\Big(p\left(y^{(e)}|x^{(e)}\right) p\left(z^{(e)}|y^{(e)}\right)\Big),\\&
\prod_{e\in\mathcal{E}} \left(\mathcal{Y}^{(e)}\times\mathcal{Z}^{(e)}\right))
\end{align*}
 and some 
 $\bar{e}\in\mathcal{E}$,  the model $\mathcal{N}_{\bar{e}}(R_c,R_p)$ replaces the degraded wiretap channel
\begin{align*} \mathcal{C}_{\bar{e}}=\big(\mathcal{X}^{(\bar{e})},p(y^{(\bar{e})}|x^{(\bar{e})})p(z^{(\bar{e})}|y^{(\bar{e})}),\mathcal{Y}^{(\bar{e})}\times\mathcal{Z}^{(\bar{e})}\big)
\end{align*}
with the noiseless degraded wiretap channel
\begin{align*} &\mathcal{C}(R_c,R_p)=(\{0,1\}^{R_c+R_p},\delta\big(y^{(\bar{e})}-(x^{(\bar{e}),c},x^{(\bar{e}),p})\big)\\&\delta\big(z^{(\bar{e})}-y^{(\bar{e}),p}\big),\{0,1\}^{R_c+R_p}\times\{0,1\}^{R_p})
\end{align*}
that delivers the rate-$R_c$ confidential portion $x^{(\bar{e}),c}$ of channel input $x^{(\bar{e})}=(x^{(\bar{e}),c},x^{(\bar{e}),p})$ to the intended receiver and the rate-$R_p$ public portion $x^{(\bar{e}),p}$ of that input to both the intended receiver and eavesdropper. The resulting network is given by
\begin{align*}& \mathcal{N}_{\bar{e}}(R_c,R_p)\displaystyle\mathop{=}^\text{def}(\{0,1\}^{R_{c}+R_{p}}\times\prod_{e\in\mathcal{E}\backslash\{\bar{e}\}}\mathcal{X}^{(e)},\\
&\delta (y^{(\bar{e})}-(x^{(\bar{e}),c},x^{(\bar{e}),p}) )\delta (z^{(\bar{e})}-y^{(\bar{e}),p} )\\
&\cdot\prod_{e\in\mathcal{E}\backslash\{\bar{e}\}} (p(y^{(e)}|x^{(e)}).p(z^{(e)}|y^{(e)}) )
,\\
&\{0,1\}^{R_{c}+R_{p}}\times\{0,1\}^{R_p}\times\prod_{e\in\mathcal{E}\backslash\{\bar{e}\}} (\mathcal{Y}^{(e)}\times\mathcal{Z}^{(e)})).
\end{align*}
\label{dfn:N_Rc_and_Rp}
\end{dfn}
 As in~\cite{network_equiv_partI, network_equiv_partII}, we allow non-integer values of $R_{c}$ and $R_{p}$ to denote noiseless bit pipes that require multiple channel uses to deliver some integer number of bits.

Many of the subsequent proofs use the notion of a ``stacked network'' introduced in~\cite{network_equiv_partI, network_equiv_partII}, extended here by adding an eavesdropper. Informally, the $N$-fold stacked network $\underline{\mathcal{N}}$ contains $N$ copies of network $\mathcal{N}$. The $N$ copies of each node $i\in\mathcal{V}$ use the outgoing messages and channel outputs from all $N$ layers of the network to form the channel inputs in each layer of the stack. Likewise, each node uses the channel outputs and messages from all layers in the stack in building its message reconstructions. An eavesdropper $E\in A$ overhears all copies of channel $e$ for each $e\in E$.

As defined formally below following~\cite{network_equiv_partI, network_equiv_partII}, a solution for $N$-fold stacked network $\underline{\mathcal{N}}$ must securely and reliably transmit, for each $i\in\mathcal{V}$ and $\mathcal{B}\in\mathcal{B}^{(i)}$, $N$ independent messages $\underline{W}^{(i\rightarrow\mathcal{B})}(1),\ldots,\underline{W}^{(i\rightarrow\mathcal{B})}(N)$ from node $i$ to all the receivers in set $\mathcal{B}$. We underline the variable names from $\mathcal{N}$ to denote variables for the stacked network $\underline{\mathcal{N}}$. Therefore $\underline{W}^{(i\rightarrow \mathcal{B})}\in\underline{\mathcal{W}}^{(i\rightarrow \mathcal{B})}\displaystyle\mathop{=}^\text{def}\left(\mathcal{W}^{(i\rightarrow \mathcal{B})}\right)^N$, $\underline{T}^{(i)}\in\underline{\mathcal{T}}^{(i)}\displaystyle\mathop{=}^{\text{def}}\big(\mathcal{T}^{(i)}\big)^N$, $\underline{X}^{(i)}_t\in\underline{\mathcal{X}}^{(i)}\displaystyle\mathop{=}^\text{def}\left(\mathcal{X}^{(i)}\right)^N$, $\underline{Y}^{(i)}_t\in\underline{\mathcal{Y}}^{(i)}\displaystyle\mathop{=}^\text{def}\left(\mathcal{Y}^{(i)}\right)^N$, and $\underline{Z}^{(e)}_t\in\underline{\mathcal{Z}}^{(e)}\displaystyle\mathop{=}^\text{def}\left(\mathcal{Z}^{(e)}\right)^N$ denote $N$-dimensional vectors of messages, channel inputs, channel outputs, and eavesdropper outputs, 
respectively, in network $\mathcal{N}$. The variables in the $\ell^\text{th}$ layer of the stack are denoted by an argument $\ell$. Finally, we define the rate $R^{(i\rightarrow\mathcal{B})}$ for a stacked network to be $(\log_2 |\underline{\mathcal{W}}^{(i\rightarrow\mathcal{B})}|)/(nN)$ since any solution of blocklength $n$ for $N$-fold stacked network $\underline{\mathcal{N}}$ can be operated as a rate-$R$ solution of blocklength $nN$ for network $\mathcal{N}$ under this definition \cite[Theorem $1$]{network_equiv_partI}. A similar argument, given in Theorem~\ref{thm:stacked_equal_nostacked} below, justifies the security constraint imposed below. 
Definitions~\ref{dfn:defining_a_network_stacked}-\ref{dfn:stacked_solution} are analogous to Definitions 4-6 in~\cite{network_equiv_partI}.
\begin{dfn}
Let a network
\begin{align*}
\mathcal{N}\displaystyle\mathop{=}^\text{def}&(\prod_{e\in\mathcal{E}}\mathcal{X}^{(e)}, \prod_{e\in\mathcal{E}}\Big(p_{e}\left(y^{(e)}|x^{(e)}\right)p_{e}\left(z^{(e)}|y^{(e)}\right)\Big),\\
& \prod_{e\in\mathcal{E}}\left(\mathcal{Y}^{(e)}\times\mathcal{Z}^{(e)}\right))
\end{align*}
be given corresponding to a graph $\mathcal{G}=(\mathcal{V}, \mathcal{E})$, and let an eavesdropper set $A\subseteq P(\mathcal{E})$ be defined on network $\mathcal{N}$. Let $\underline{\mathcal{N}}$ be the $N$-fold stacked network for $\mathcal{N}$. A blocklength-$n$ solution $\mathcal{S}(\underline{\mathcal{N}})$ to this network is defined as a set of encoding functions
\begin{align*}
\underline{X}^{(i)}_t&:\left(\underline{\mathcal{Y}}^{(i)}\right)^{t-1}\times \prod_{\mathcal{B}\in\mathcal{B}^{(i)}}\underline{\mathcal{W}}^{(i\rightarrow\mathcal{B})}\times\underline{\mathcal{T}}^{(i)}\longrightarrow \underline{\mathcal{X}}^{(i)}
\end{align*}
mapping $\left(\underline{Y}^{(i)}_1,\ldots, \underline{Y}^{(i)}_{t-1}, \big(\underline{W}^{(i\rightarrow \mathcal{B})}:\mathcal{B}\in\mathcal{B}^{(i)}\big),\underline{T}^{(i)}\right)$ to $\underline{X}^{(i)}_t$ for each $i\in\mathcal{V}$ and $t\in\{1,\ldots,n\}$,
and  decoding functions
\begin{align*}
\underline{\breve{W}}^{(j\rightarrow\mathcal{K}, i)}&:\left(\underline{\mathcal{Y}}^{(i)}\right)^n\times \prod_{\mathcal{B}\in\mathcal{B}^{(i)}}\underline{\mathcal{W}}^{(i\rightarrow\mathcal{B})}\times\underline{\mathcal{T}}^{(i)}\longrightarrow \underline{\mathcal{W}}^{(j\rightarrow\mathcal{K})}
\end{align*}
 mapping $\left(\underline{Y}^{(i)}_1,\ldots, \underline{Y}^{(i)}_n,\big(\underline{W}^{(i\rightarrow \mathcal{B})}:\mathcal{B}\in\mathcal{B}^{(i)}\big),\underline{T}^{(i)}\right)$ to $\underline{\breve{W}}^{(j\rightarrow\mathcal{K}, i)}$ for each $j\in\mathcal{V}$, $\mathcal{K}\in\mathcal{B}^{(j)}$, and $i\in\mathcal{K}$. The solution $\mathcal{S}(\underline{\mathcal{N}})$ is called a $(\lambda, \varepsilon,A, R)$--solution for stacked network $\underline{\mathcal{N}}$, denoted $(\lambda, \varepsilon,A,R)$--$\mathcal{S}(\underline{\mathcal{N}})$, if $\left(\log_2\left|\underline{W}^{(i\rightarrow \mathcal{B})}\right|\right)\slash(nN)= R^{(i\rightarrow\mathcal{B})}$, $I\left(\left(\underline{Z}^{(E)}\right)^n;\underline{W}\right)<nN\varepsilon$ for every $E\in A$, and $\Pr\left(\underline{\breve{W}}^{(j\rightarrow\mathcal{K}, i)}\neq \underline{W}^{(j\rightarrow\mathcal{K})}\right)<\lambda$ for  the specified encoding and decoding functions.
\label{dfn:defining_a_network_stacked}
\end{dfn}

\begin{dfn}
The $A$-secure rate region $\mathcal{R}(\underline{\mathcal{N}},A)\subseteq \mathbb{R}\hspace{0.4mm}^{m(2^{m-1}-1)}_+$ of stacked network $\underline{\mathcal{N}}$ is the closure of all rate vectors $R$ such that for any $\lambda>0$ and any $\varepsilon>0$, a solution $(\lambda,\varepsilon,A, R)$--$\mathcal{S}(\underline{\mathcal{N}})$ exists for sufficiently large $N$.
\label{dfn:defining_a_rate_region_stacked}
\end{dfn}

%
\begin{dfn}
Let a network
\begin{align*}
\mathcal{N}\displaystyle\mathop{=}^\text{def}&(\prod_{e\in\mathcal{E}}\mathcal{X}^{(e)}, \prod_{e\in\mathcal{E}}\Big(p_{e}\left(y^{(e)}|x^{(e)}\right)p_{e}\left(z^{(e)}|y^{(e)}\right)\Big),\\
& \prod_{e\in\mathcal{E}} \mathcal{Y}^{(e)}\times\prod_{e\in\mathcal{E}}\mathcal{Z}^{(e)})
\end{align*}
be given corresponding to a graph $\mathcal{G}=(\mathcal{V}, \mathcal{E})$. Fix positive integers $n$ and $N$  as the blocklength and stack size, respectively. For each $i\in\mathcal{V}$ and $\mathcal{B}\in\mathcal{B}^{(i)}$, let $R^{(i\rightarrow\mathcal{B})}$ and $\tilde{R}^{(i\rightarrow\mathcal{B})}$ be constants with $\tilde{R}^{(i\rightarrow\mathcal{B})}\geq R^{(i\rightarrow\mathcal{B})}$. Define $W^{(i\rightarrow\mathcal{B})}=\{1,\ldots,2^{nR^{(i\rightarrow\mathcal{B})}}\}$ and $\tilde{W}^{(i\rightarrow\mathcal{B})}=\{1,\ldots,2^{n\tilde{R}^{(i\rightarrow\mathcal{B})}}\}$. Let $\underline{\mathcal{N}}$ be the $N$-fold stacked network for $\mathcal{N}$. A blocklength-$n$ stacked solution $\underline{\mathcal{S}}(\underline{\mathcal{N}})$ to this network is defined as a set of mappings
\begin{align*}
\underline{\tilde{W}}^{\text{\raisebox{-0.6mm}{$(i\rightarrow\mathcal{B})$}}}: &\underline{\mathcal{W}}^{(i\rightarrow\mathcal{B})}\rightarrow \underline{\tilde{\mathcal{W}}}^{\text{\raisebox{-0.6mm}{$(i\rightarrow\mathcal{B})$}}}\\
X_t^{(i)}: &\left(\mathcal{Y}^{(i)}\right)^{t-1}\times \prod_{\mathcal{B}\in\mathcal{B}^{(i)}}\mathcal{\tilde{W}}^{(i\rightarrow\mathcal{B})}\times\mathcal{T}^{(i)}\longrightarrow \mathcal{X}^{(i)}\\
\breve{\tilde{W}}^{(j\rightarrow\mathcal{K}, i)}: &\left(\mathcal{Y}^{(i)}\right)^n\times \prod_{\mathcal{B}\in\mathcal{B}^{(i)}}\mathcal{\tilde{W}}^{(i\rightarrow\mathcal{B})}\times\mathcal{T}^{(i)}\longrightarrow \mathcal{\tilde{W}}^{(j\rightarrow\mathcal{K})}\\
\underline{\breve{W}}^{\text{\raisebox{-0.9mm}{$(j\rightarrow\mathcal{K}, i)$}}}: &\underline{\tilde{\mathcal{W}}}^{\text{\raisebox{-0.9mm}{$(j\rightarrow\mathcal{K})$}}}\rightarrow \underline{\mathcal{W}}^{(j\rightarrow\mathcal{K})},
\end{align*}
where the other channel encoder $\underline{\tilde{W}}^{\text{\raisebox{-0.6mm}{$(i\rightarrow\mathcal{B})$}}}(\cdotp)$ encodes message $\underline{W}^{(i\rightarrow\mathcal{B})}$ to $\underline{\tilde{W}}^{\text{\raisebox{-0.6mm}{$(i\rightarrow\mathcal{B})$}}}\big(\underline{W}^{(i\rightarrow\mathcal{B})}\big)$, encoder $X_t^{(i)}(\cdotp)$ independently encodes each dimension $\ell\in\{1,\ldots,N\}$ of outgoing messages $\underline{\tilde{W}}^{\text{\raisebox{-0.6mm}{$(i\rightarrow\mathcal{B})$}}}$, received channel outputs $\underline{Y}_1^{(i)},\ldots,\underline{Y}_{t-1}^{(i)}$, and random keys $\underline{T}^{(i)}$ to channel input $$X^{(i)}_t(\underline{Y}^{(i)}_1(\ell),\ldots,\underline{Y}^{(i)}_{t-1}(\ell),
 \big(\underline{\tilde{W}}^{\text{\raisebox{-0.6mm}{$(i\rightarrow\mathcal{B})$}}}(\ell):\mathcal{B}\in\mathcal{B}^{(i)}\big),\underline{T}^{(i)}(\ell)),$$ node decoder $\breve{\tilde{W}}^{(j\rightarrow\mathcal{K},i)}(\cdotp)$ independently decodes each dimension of the reconstruction $$\breve{\tilde{W}}^{(j\rightarrow\mathcal{K},i)}(\underline{Y}^{(i)}_1(\ell),\ldots,\underline{Y}^{(i)}_n(\ell),\\
 \big(\underline{\tilde{W}}^{\text{\raisebox{-0.6mm}{$(i\rightarrow\mathcal{B})$}}}(\ell):\mathcal{B}\in\mathcal{B}^{(i)}\big),\underline{T}^{(i)}(\ell))$$ of $\underline{\tilde{W}}^{\text{\raisebox{-0.6mm}{$(j\rightarrow\mathcal{K})$}}}$ at node $i$, and channel decoder $\underline{\breve{W}}^{\text{\raisebox{-0.9mm}{$(j\rightarrow\mathcal{K}, i)$}}}(\cdotp)$ reconstructs message vector $ \underline{\breve{W}}^{\text{\raisebox{-0.9mm}{$(j\rightarrow\mathcal{K}, i)$}}}(\underline{\breve{\tilde{W}}}^{\text{\raisebox{-1.8mm}{$(j\rightarrow\mathcal{K},i)$}}})$.
\label{dfn:stacked_solution}
\end{dfn}

The following theorem extends~\cite[Theorem $2$]{network_equiv_partI} from traditional to secure capacity.
\begin{theorem}
The rate regions $\mathcal{R}(\mathcal{N},A)$ and $\mathcal{R}(\underline{\mathcal{N}},A)$ are identical. Further, for any $R\in\text{int}\big(\mathcal{R}(\mathcal{N},A)\big)$, there exists a sequence of $(2^{-N\delta}, \varepsilon,A, R)$--$\underline{\mathcal{S}}(\underline{\mathcal{N}})$ stacked solutions for the stacked network $\mathcal{\underline{N}}$ for some $\delta>0$.
\label{thm:stacked_equal_nostacked}
\end{theorem}
%

\noindent {\it Sketch of the proof:}
The argument to show $\mathcal{R}(\underline{\mathcal{N}}, A)\subseteq \mathcal{R}(\mathcal{N}, A)$ follows~\cite[Theorem $1$]{network_equiv_partI}: given any $R\in\text{int}(\mathcal{R}(\underline{\mathcal{N}},A))$, a blocklength-$n$ $(\lambda,\varepsilon,A, R)-\mathcal{S}(\underline{\mathcal{N}})$ solution for network $\underline{\mathcal{N}}$ is unraveled across time to achieve a blocklength-$nN$ solution for network $\mathcal{N}$. Since the given code satisfies the causality constraints and precisely implements the operations of $\mathcal{S}(\underline{\mathcal{N}})$, the solution $\mathcal{S}(\mathcal{N})$ achieves the same rate, error probability, and secrecy on $\mathcal{N}$ as the solution $\mathcal{S}(\underline{\mathcal{N}})$ achieves on $\underline{\mathcal{N}}$, which gives the forward result.

The converse  follows~\cite[Theorem $2$]{network_equiv_partI}. Again, fix $\varepsilon>0$, and for any $R\in\text{int}\big(\mathcal{R}(\mathcal{N},A)\big)$ choose $\tilde{R}\in\text{int}\big(\mathcal{R}(\mathcal{N},A)\big)$ with $\tilde{R}^{(i\rightarrow\mathcal{B})}>R^{(i\rightarrow\mathcal{B})}$ for all $(i,\mathcal{B})$ with $R^{(i\rightarrow\mathcal{B})}>0$. Define $\rho = \min_{i\in\mathcal{V}}\min_{\mathcal{B}\in\mathcal{B}^{(i)}}\big(\tilde{R}^{(i\rightarrow\mathcal{B})}-R^{(i\rightarrow\mathcal{B})}\big)$ and choose constant $\lambda>0$ satisfying
\begin{align*}
\max_{i\in\mathcal{V}}\max_{\mathcal{B}\in\mathcal{B}^{(i)}}\tilde{R}^{(i\rightarrow\mathcal{B})}\lambda+h(\lambda)<\rho.
\end{align*}
This is possible by choosing $\lambda$ small enough so that $\lambda<\rho\slash(3\max_{i\in\mathcal{V}}\max_{\mathcal{B}\in\mathcal{B}^{(i)}}\tilde{R}^{(i\rightarrow\mathcal{B})})$ and $h(\lambda)<\rho\slash(3\rho)$. Since $\tilde{R}^{(i\rightarrow\mathcal{B})}>R^{(i\rightarrow\mathcal{B})}$, there exists a blocklength $n$ such that a  $(\lambda,\frac{\varepsilon}{3},A,\tilde{R})$--$\mathcal{S}(\mathcal{N})$ single-layer solution exists. A stacked solution is built using this same $(\lambda,\frac{\varepsilon}{3},A,R)$--$\mathcal{S}(\mathcal{N})$ single-layer solution in each layer and a randomly chosen channel code across the layers of the stack.
$\hfill\Box$

\section{Main Results}
\label{sec:intuition}

In Theorem~\ref{thm:upper_bound}, we show that for any network $\mathcal{N}$ of wiretap channels and any edge $\bar{e}\in\mathcal{E}$, replacing channel $\mathcal{C}_{\bar{e}}$ with a noiseless degraded wiretap channel 
of appropriate capacities $R_c$ and $R_p$, as shown in Figure~\ref{fig:rate_regions_noiseless}, yields a  network $\mathcal{N}_{\bar{e}}(R_c,R_p)$ (Definition~\ref{dfn:N_Rc_and_Rp}) whose  secure capacity region contains the secure capacity region of $\mathcal{N}$. Theorem~\ref{thm:upper_bound}  extends~\cite[Theorem $5$]{network_equiv_partII} from traditional to secure capacity.

\begin{theorem}
Consider a network $\mathcal{N}$ and an adversarial set $A\subseteq\mathcal{P}(\mathcal{E})$. $\mathcal{R}(\mathcal{N},A)\subseteq\mathcal{R}(\mathcal{N}_{\bar{e}}(R_c,R_p),A)$ for
\begin{align*}
R_c &> \mathop{\max}_{p(x^{(\bar{e})})}I(X^{(\bar{e})};Y^{(\bar{e})}) -\mathop{\max}_{p(x^{(\bar{e})})}I(X^{(\bar{e})};Z^{(\bar{e})})\\
R_p &> \mathop{\max}_{p(x^{(\bar{e})})}I(X^{(\bar{e})};Z^{(\bar{e})}).
\end{align*}
\label{thm:upper_bound}
\end{theorem}
\noindent {\it Sketch of the proof:}
By Theorem~\ref{thm:stacked_equal_nostacked} it suffices to prove $\mathcal{R}(\underline{\mathcal{N}},A)\subseteq \mathcal{R}(\underline{\mathcal{N}}_{\bar{e}}(R_c,R_p),A)$. We employ a channel code across layers of the stack to emulate a secure code for network $\underline{\mathcal{N}}$ on network $\underline{\mathcal{N}}_{\bar{e}}(R_c,R_p)$.  Typical inputs $\underline{X}_t$ to $\bar{e}$ are mapped to jointly typical outputs from a random codebook.  It can be shown that the induced probability distribution $p'$ is close to the probability distribution $p$ of the original secure code for $\underline{\mathcal{N}}$, and that mutual information values under both probability distributions are similar. The bits transmitted over the noiseless channel correspond to the codeword index, and thus reveal a similar amount of information to the wiretapper as its observations of the original noisy channel. 
%
$\hfill\Box$

Theorem~\ref{thm:lower_bound_one_or_no_eavesdropped} shows cases where  the upper bound shown in Theorem~\ref{thm:upper_bound} is tight.
\begin{theorem}
Consider a network $\mathcal{N}$, an adversarial set $A\subseteq\mathcal{P}(\mathcal{E})$, and a single link $\bar{e}\in\mathcal{E}$. Let
\begin{align*}
R_c &= \mathop{\max}_{p(x^{(\bar{e})})}I(X^{(\bar{e})};Y^{(\bar{e})}) -\mathop{\max}_{p(x^{(\bar{e})})}I(X^{(\bar{e})};Z^{(\bar{e})})\\
R_p &= \mathop{\max}_{p(x^{(\bar{e})})}I(X^{(\bar{e})};Z^{(\bar{e})}).
\end{align*}
If $\bar{e}$ is invulnerable to wiretapping ($\bar{e}\notin E$ for all $E\in A$) or is not simultaneously wiretapped with other links ($\bar{e}\in E$ implies $|E|=1$), then $\mathcal{R}(\mathcal{N},A)=\mathcal{R}(\mathcal{N}_{\bar{e}}(R_c,R_p),A)$.
\label{thm:lower_bound_one_or_no_eavesdropped}
\end{theorem}

\begin{figure*}
\centerline{\includegraphics[width=1.7\columnwidth]{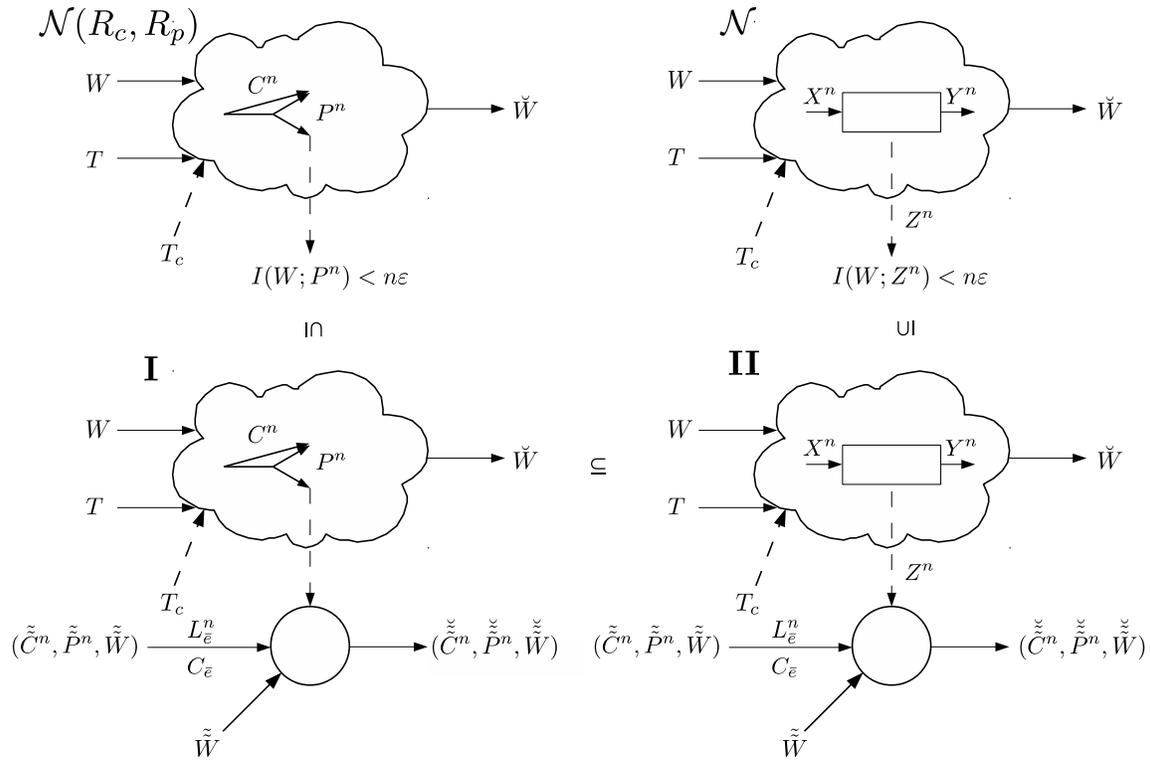}}
\caption{Network $\mathcal{N}_{\bar{e}}(R_c,R_p)$ along with networks \textbf{I}, \textbf{II} and $\mathcal{N}$ that assist proving Theorem~\ref{thm:lower_bound_one_or_no_eavesdropped}. }
\label{fig:lower_bound_single}
\end{figure*}
\noindent {\it Sketch of the proof:}
We outline the proof for the case where $\bar{e}$ is wiretapped but not simultaneously with other links; the case where it is invulnerable to wiretapping is a simpler version.

We first show that  $\mathcal{R}(\mathcal{N}_{\bar{e}}(R_c-\epsilon,R_p-\epsilon),A) \subseteq\mathcal{R}(\mathcal{N},A)$ for any $\epsilon>0$, by starting with a secure code of rate $R$ for network $\mathcal{N}_{\bar{e}}(R_c,R_p)$ and constructing a corresponding secure code for network $\mathcal{N}$.
Denote by $C_t$ and $P_t$ the 
transmissions across the confidential and public links, respectively, of edge $\bar{e}\in\mathcal{E}$ at time $t$. Let $C^n=(C_1,\ldots,C_n)$, $P^n=(P_1,\ldots,P_n)$ and denote by $C_j^i$ and $P_j^i$ for any $j<i$ the vectors $C_j^i=(C_j,C_{j+1},\ldots,C_i)$ and $P_j^i=(P_j,P_{j+1},\ldots,P_i)$. 
We define networks \textbf{I} and \textbf{II} shown in Figure~\ref{fig:lower_bound_single} that are identical to networks $\mathcal{N}_{\bar{e}}(R_c,R_p)$ and $\mathcal{N}$ respectively with the addition of an auxiliary receiver that observes the wiretap output of $\bar{e}$, messages $W$ and a noiseless side channel of capacity $C_{\bar{e}}$ (defined below) from a ``super-source'' that has access to $(W,C^n,P^n)$. In network \textbf{I} (\textbf{II}) the auxiliary receiver is required to decode the confidential bits $C^n$. 

We construct a code for a stacked version of network \textbf{I} with $N_1$ layers in which the auxiliary receiver is   able to decode the confidential bits $C^n$. 
The constructed coded for the stacked version of network \textbf{I} can be seen as a code of blocklength $n_1 = nN_1$ for the non-stacked version of network \textbf{I}. To move the proof from network \textbf{I} to network \textbf{II} we use a stacked version of network \textbf{II} with $N_2$ layers. 
The code used at each layer of the stacked version of network \textbf{II} is the code of blocklength $n_1$ constructed above. We need to use a stacked version of network \textbf{II} to use a channel code at edge $\bar{e}$ of network \textbf{II} to emulate the noiseless edge $\bar{e}$ of network \textbf{I}.

In the following we show that the communication code of network \textbf{II} gives a secure code of network $\mathcal{N}$.  These auxiliary  receivers assist in the proof of the secrecy of the code for the eavesdropping set $\{e\}\in A$ in the following manner: capacity $C_{\bar{e}}$ is defined such that the sum of capacities of $(W,Z^n,L_{\bar{e}}^n)$ (where $L_{\bar{e}}^n$ are the bits in the noiseless bit pipe of capacity$C_{\bar{e}}$) that are all the incoming links to the auxiliary receivers is almost equal to the entropy of $(P^n, C^n, W)$ that correspond to the decoded message at the auxiliary receivers and therefore all links are filled up to capacity. Therefore there is no spare capacity at links $Z^n$ to carry any information about message $W$ and therefore the code is secure.

On the other hand, the upper bound result in Theorem~\ref{thm:upper_bound} implies that  $\mathcal{R}(\mathcal{N},A) \subseteq\mathcal{R}(\mathcal{N}_{\bar{e}}(R_c+\epsilon,R_p+\epsilon),A)$ for any $\epsilon>0$.
We then prove   a continuity result on the rate region $\mathcal{R}(\mathcal{N}_{\bar{e}}(R_{c},R_{p}),A)$ with respect to $(R_c,R_p)$ when $R_c>0$ and $R_p>0$. The lower bound result, the upper bound result, and the continuity result  together prove Theorem~\ref{thm:lower_bound_one_or_no_eavesdropped}.
$\hfill\Box$

Example~\ref{exam:one_and_only} demonstrates  applications of Theorem~\ref{thm:upper_bound} and~\ref{thm:lower_bound_one_or_no_eavesdropped} and shows that while Theorem~\ref{thm:upper_bound} is tight  in many cases, it is not always tight when the replaced link appears in one or more eavesdropping sets of size greater than $1$. 

\begin{figure*}
\centering
\subfigure[]{\label{fig:super_network_noisy}\includegraphics[width=0.85\columnwidth]{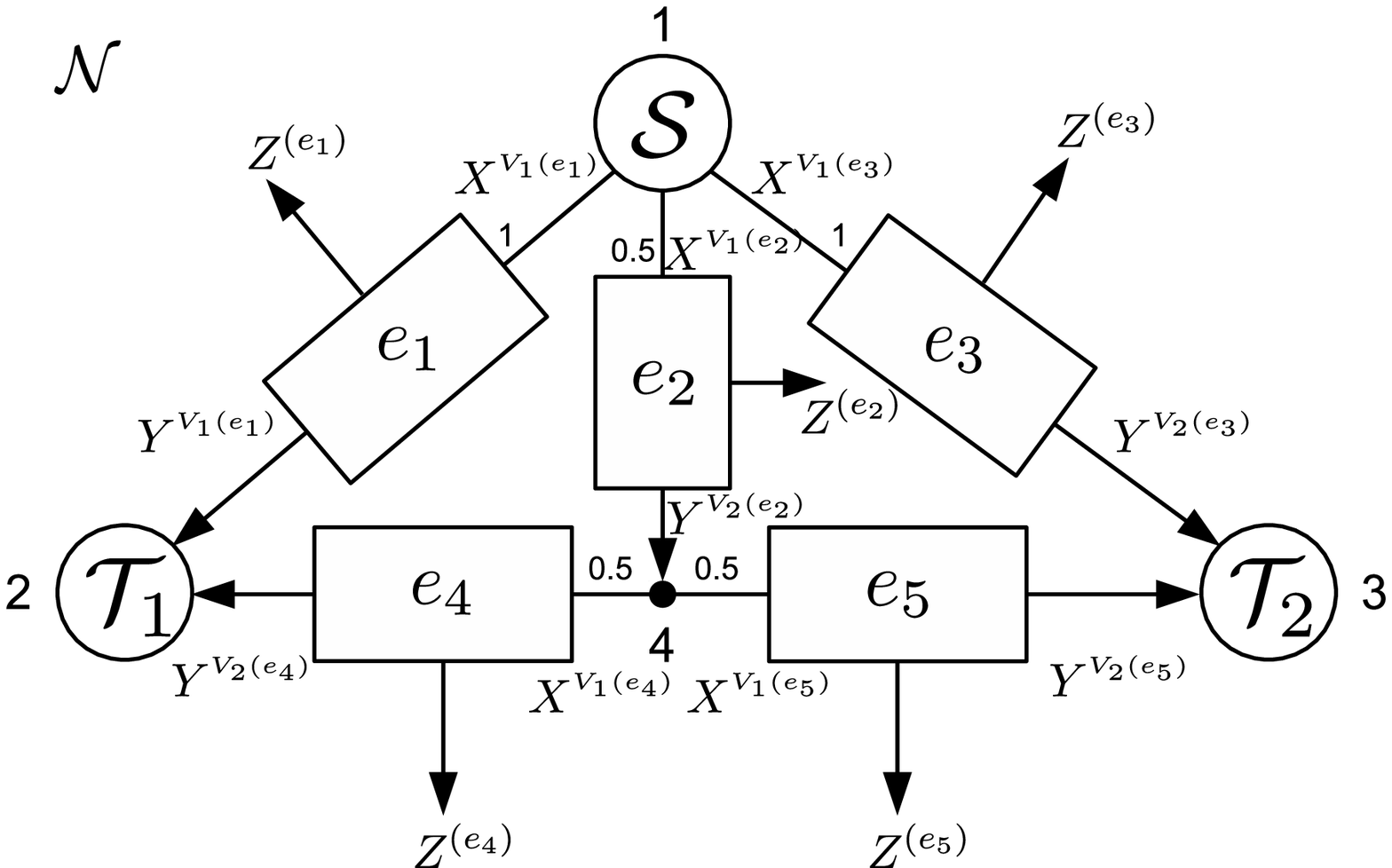}}
\subfigure[]{\label{fig:super_network_intermediate}\includegraphics[width=0.85\columnwidth]{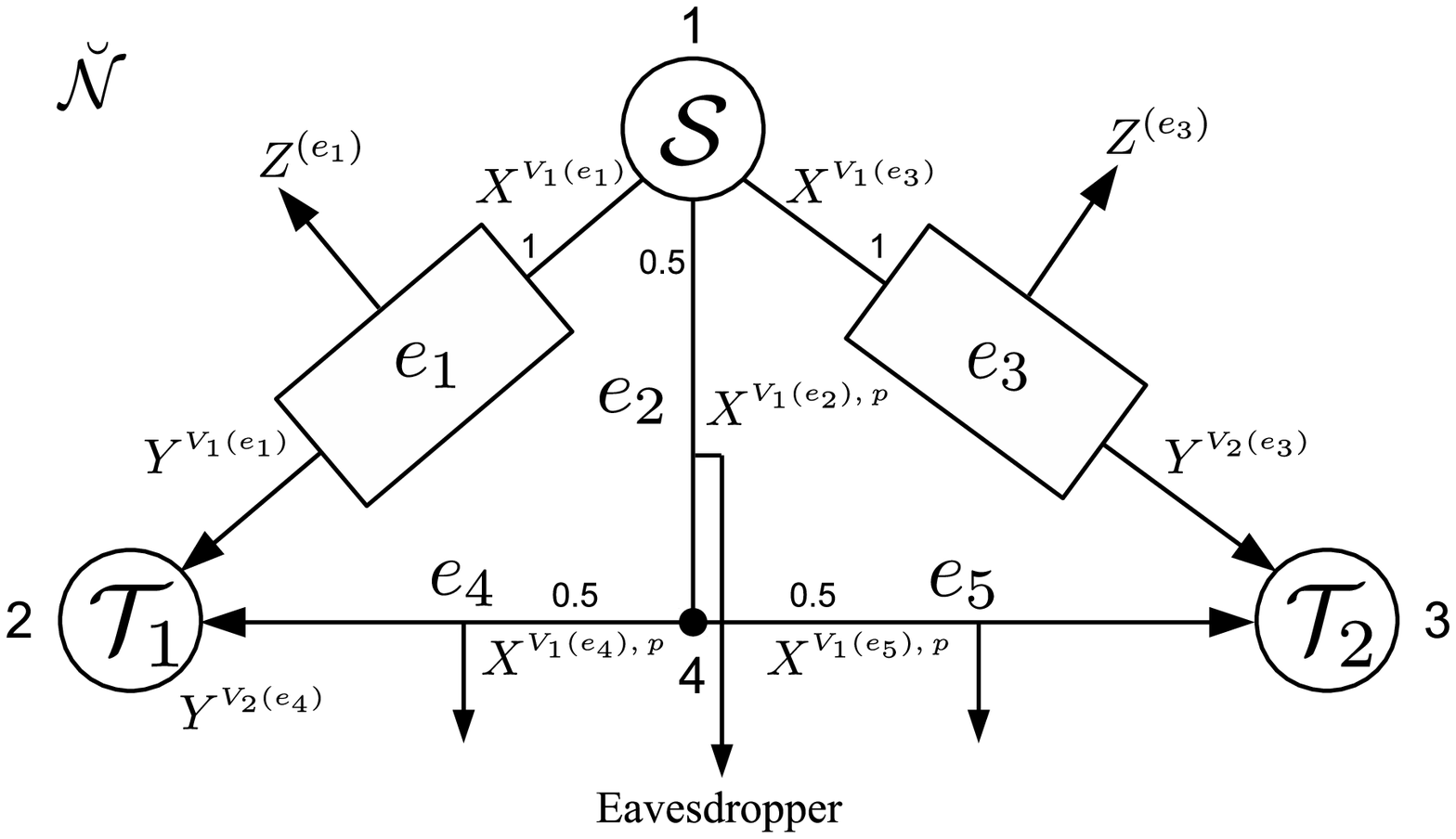}}
\subfigure[]{\label{fig:super_network_noiseless}\includegraphics[width=0.85\columnwidth]{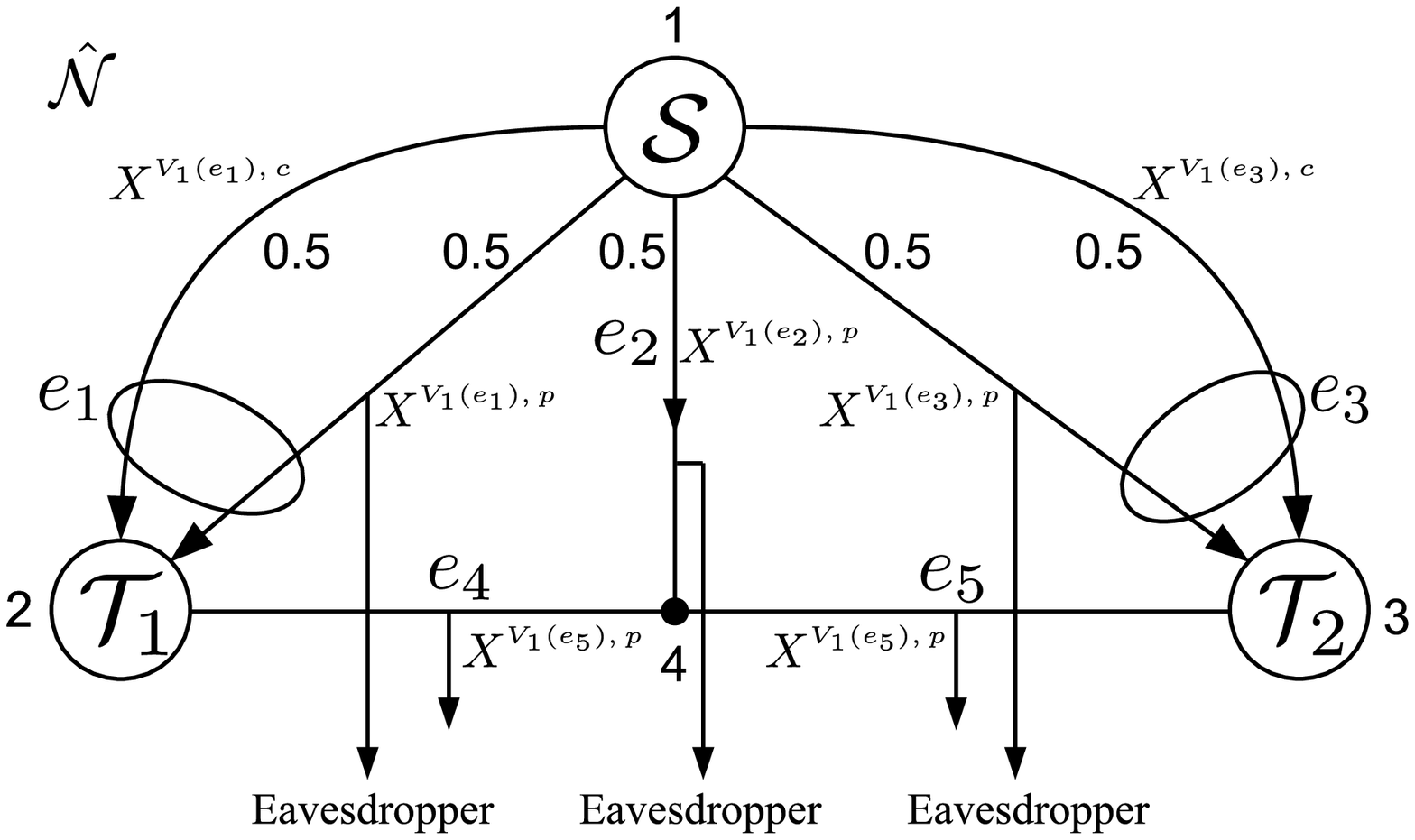}}
\subfigure[]{\label{fig:super_network_code}\includegraphics[width=0.85\columnwidth]{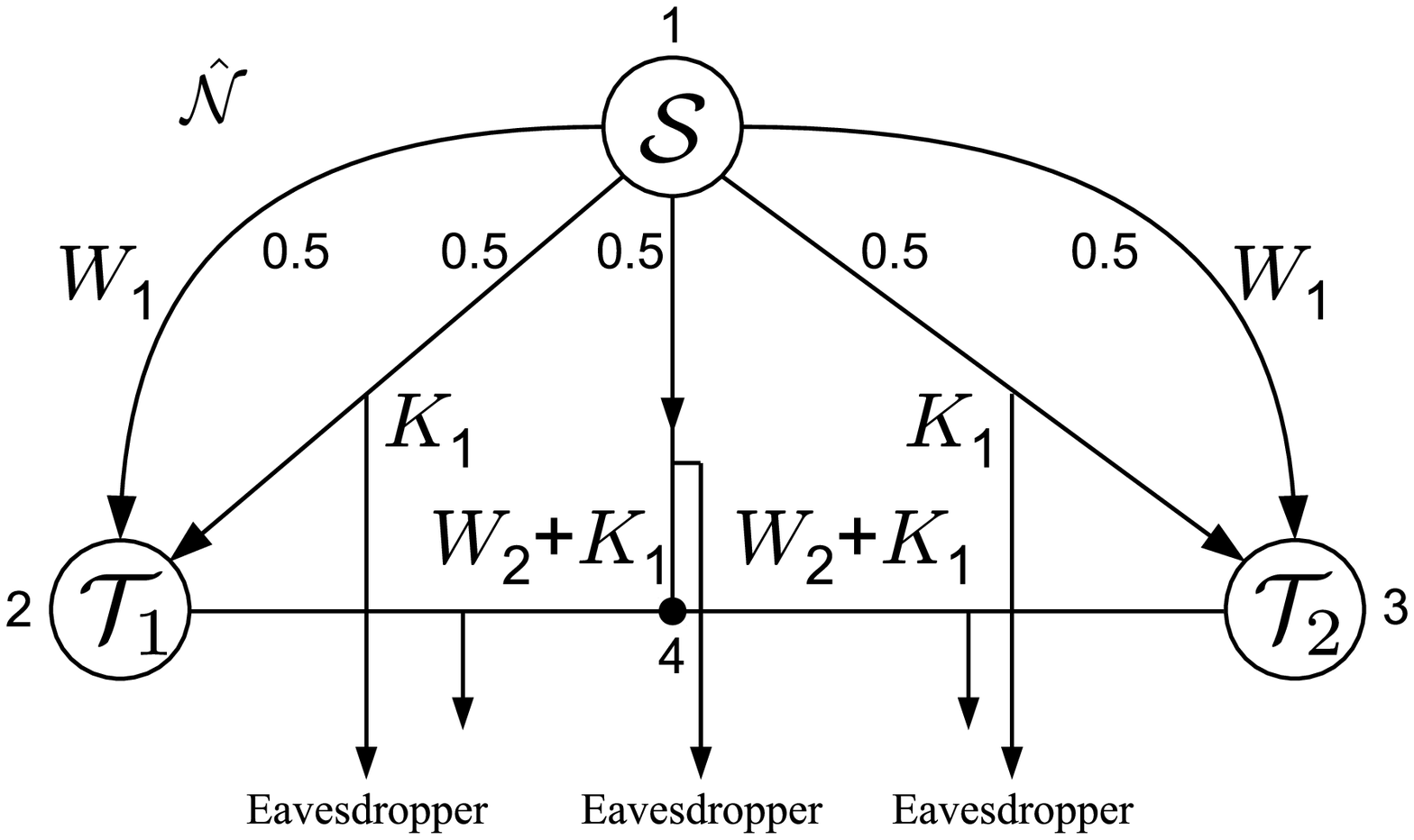}}
\subfigure[]{\label{fig:super_network_channel_e1}\includegraphics[width=0.85\columnwidth]{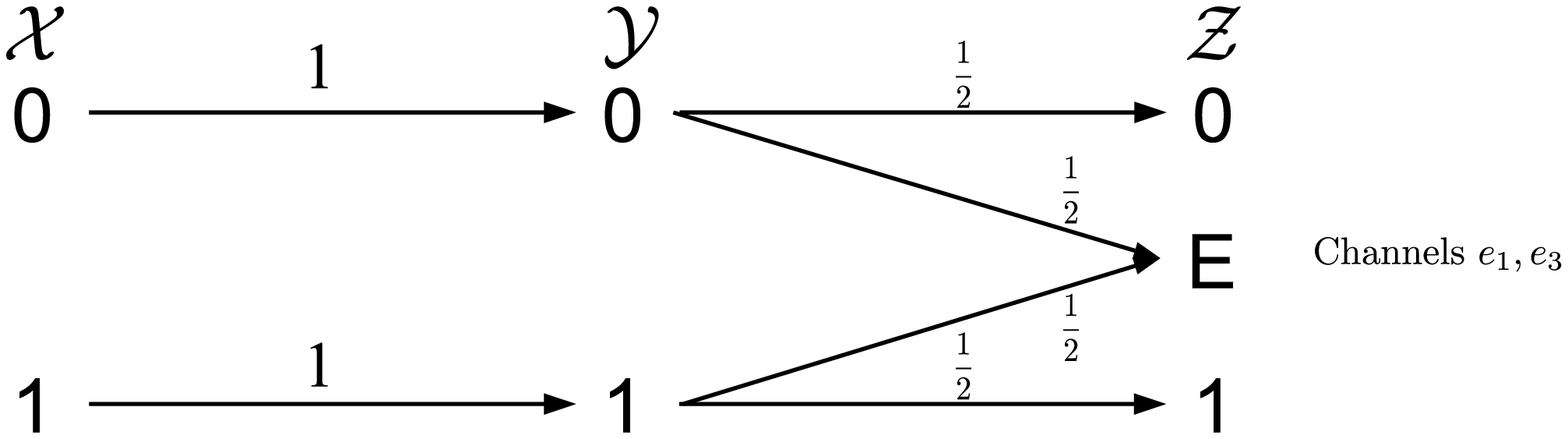}}
\subfigure[]{\label{fig:super_network_channel_e2}\includegraphics[width=0.85\columnwidth]{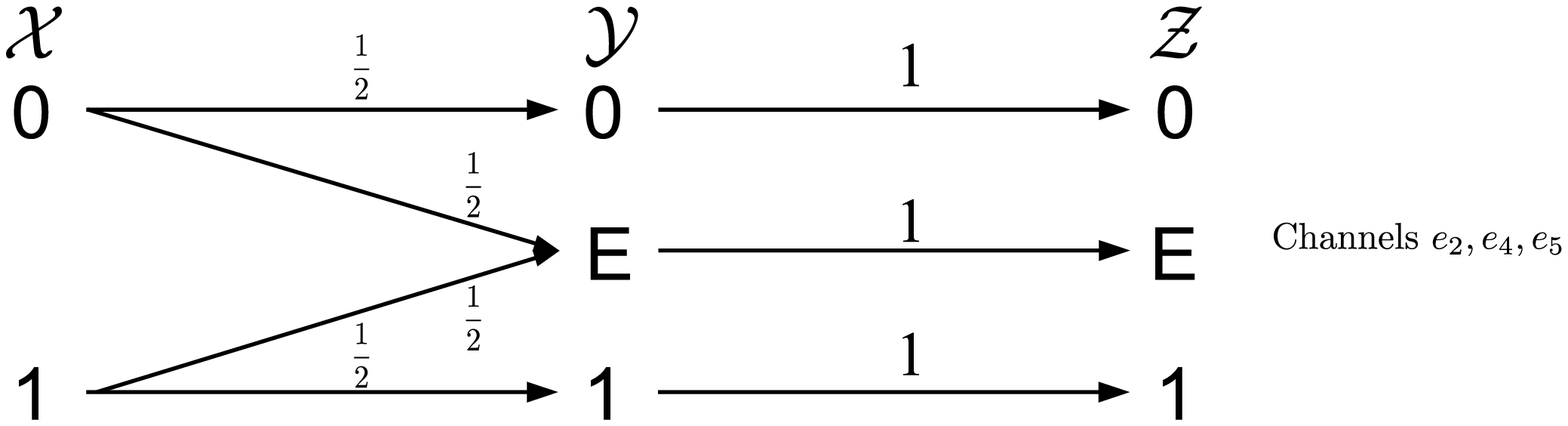}}
\caption{(a) The network for Example~\ref{exam:one_and_only} and (b) its equivalent model by replacing channels $e_2$, $e_4$, and $e_5$ by their equivalent noiseless links by Theorem~\ref{thm:lower_bound_one_or_no_eavesdropped} (rate-$0$ links are omitted from the model). (c) The noiseless model of (a) by applying Theorem~\ref{thm:upper_bound} and (d) the secrecy capacity achieving code for the network in (c). (e), (f) The channel distributions for independent degraded wiretap channels $e_1$, $e_3$ and $e_2$, $e_4$, $e_5$ respectively.}
\label{fig:super_network}
\end{figure*}
\begin{exam}
In the network of Figure~\ref{fig:super_network_noisy}, channels $e_1=(1,2,1)$, $e_2=(1,4,1)$, $e_3=(1,3,1)$, $e_4=(4,2,1)$, and $e_5=(4,3,1)$  are independent degraded binary wiretap channels. Channels $e_1$ and $e_3$ have erasure probability $0$ at each intended receiver and erasure probability $\frac{1}{2}$ at each wiretap output, as shown in Figure~\ref{fig:super_network_channel_e1}. Channels $e_2$, $e_4$, and $e_5$ have erasure probability $\frac{1}{2}$, with identical outputs for their intended and eavesdropped outputs, as shown in Figure~\ref{fig:super_network_channel_e2}. We consider a single multicast from source $\mathcal{S}$ at node $1$ to terminals $\mathcal{T}_1$ and $\mathcal{T}_2$ at nodes $2$ and $3$. We therefore set $R^{(i\rightarrow\mathcal{B})}=0$ for all $(i,\mathcal{B})\neq(1,\{2,3\})$ and then consider the point $R\in\mathcal{R}(\mathcal{N},A)$ that maximizes $R^{(1\rightarrow\{2,3\})}$. The eavesdropper can listen in on either both $e_1$ and $e_3$ or just $e_2$, {\it i.e.}, $A = \big\{\{e_1,e_3\}, \{e_2\}\big\}$.  The network $\breve{\mathcal{N}}$ shown in Figure~\ref{fig:super_network_intermediate} has secrecy capacity under adversarial set $A = \big\{\{e_1,e_3\}, \{e_2\}\big\}$ identical to that of the network in Figure~\ref{fig:super_network_noisy} $\big(\mathcal{R}(\mathcal{N},A)=\mathcal{R}(\breve{\mathcal{N}},A)\big)$ and is obtained by three applications of Theorem~\ref{thm:upper_bound}. Here channel $\mathcal{C}_{e_4}$ and $\mathcal{C}_{e_5}$ have been replaced by channel $\mathcal{C}(\frac{1}{2},0)$ since channels $e_4$ and $e_5$ are invulnerable to eavesdropping ($e_4, e_5\notin E$ for all $E\in A$). Likewise $\mathcal{C}_{e_2}$ has been replaced by $\mathcal{C}(0,\frac{1}{2})$ since $e_2$ cannot be simultaneously eavesdropped with any other channel ($e_2\in E$ implies $|E|=1$) and has $0$ confidential bits. The noiseless network $\hat{\mathcal{N}}$ is an upper bounding model for the network in Figure~\ref{fig:super_network_intermediate} (and therefore also an upper bounding model for the network in Figure~\ref{fig:super_network_noisy}, giving $\mathcal{R}(\mathcal{N},A)=\mathcal{R}(\breve{\mathcal{N}},A)\subseteq\mathcal{R}(\hat{\mathcal{N}},A)$), and is obtained by two applications of Theorem~\ref{thm:upper_bound}, replacing channels $e_1$ and $e_3$ by their upper bounding models.

A rate-$1$ blocklength-$2$ code for network $\hat{\mathcal{N}}$ is shown in Figure~\ref{fig:super_network_code}. The message $W^{(1\rightarrow\{2,3\})}\in\{0,1\}^2$ is broken into a pair of messages $W^{(1\rightarrow\{2,3\})} = \big(W_1,W_2\big)\in\{0,1\}^2$ with $H\big(W_1\big) = H\big(W_2\big) = 1$ and $H\big(W_1,W_2\big) = 2$. Random key $K_1\in\{0,1\}$ is chosen uniformly at random and independently of $\big(W_1,W_2\big)$. The code is secure since $I\big(W_1,W_2;K_1\big) = 0$ and $I\big(W_1,W_2;W_2+K_1\big)=0$. 
In \cite{eav_equvpreprint} we prove using information inequalities that the noisy network $\mathcal{N}$ of Figure~\ref{fig:super_network_noisy} has multicast secrecy capacity at most $0.875$.
\label{exam:one_and_only}
\end{exam}

To provide some intuition, notice that our capacity-achieving code for $\hat{\mathcal{N}}$ transmits the same key over a pair of noiseless links ($e_1$ and $e_3$ in $\hat{\mathcal{N}}$). Direct emulation of this solution 
in $\breve{\mathcal{N}}$ network in Figure~\ref{fig:super_network_noisy} fails to maintain security. Specifically, if the same input is transmitted over channels $e_1$ and $e_3$ ($X^{(e_1)}_t=X^{(e_3)}_t$ for all $t\in\{1,\ldots,n\}$), then an eavesdropper accessing $E=\{e_1,e_3\}$ sees independent channel outputs $Z^{(e_1)}_t$ and $Z^{(e_3)}_t$ resulting from the same channel input $X^{(e_1)}_t=X^{(e_3)}_t$ at each time $t$. Since each transmitted bit is erased with probability $\frac{1}{2}$ and the erasure events are independent by assumption, an eavesdropper that wiretaps both $e_1$ and $e_3$ is expected to receive roughly $75\%$ of the transmitted information bits. Consequently, a key of rate $0.5$ is not enough to completely protect $W^{(1\rightarrow\{2,3\})}$ from the eavesdropper in this case. 
The problem here is that transmitting correlated information on multiple channels may be necessary to achieve the secure capacity in the noiseless case, but the same strategy may fail in the noisy case owing to independent realizations of probabilistic noise on different channels. 

Theorems~\ref{thm:arbitrary_eavesdropped_lower_bound_noiseless_to_noisy} and~\ref{thm:arbitrary_eavesdropped_lower_bound_noiseless_to_noisy_model_2} provide two different lower bounds for the case of multiple wiretapped channels. These bounds correspond to achievable schemes that ensure all links to the eavesdropper are filled to capacity with independent randomness.

\subsubsection*{Lower bound model-I}
\label{subsecsec:Model_p2p_noisy_intuition}

The first lower bound results from removing the public portion of the upper bounding model. The lower bound is achievable since it is always possible to simply avoid the transmission of any rate on channel $\bar{e}$ that can be overheard by the eavesdropper.
\begin{theorem}
Consider a network $\mathcal{N}$, an adversarial set $A\subseteq\mathcal{P}(\mathcal{E})$, and a single link $\bar{e}\in\mathcal{E}$.  $\mathcal{R}(\mathcal{N}_{\bar{e}}(R_c,0),A)\subseteq\mathcal{R}(\mathcal{N},A)$ for
\begin{align*}
R_c &< \mathop{\max}_{p(x^{(\bar{e})})}I(X^{(\bar{e})};Y^{(\bar{e})})-\mathop{\max}_{p(x^{(\bar{e})})}I(X^{(\bar{e})};Z^{(\bar{e})}).
\end{align*}

\label{thm:arbitrary_eavesdropped_lower_bound_noiseless_to_noisy}
\end{theorem}
\noindent{\it Sketch of the proof:}
The proof of this theorem is similar to the proof of Theorem~\ref{thm:lower_bound_one_or_no_eavesdropped} except that in the noisy network we transmit independent random bits in place of public bits.
$\hfill\Box$

The lower bound model-I of Theorem~\ref{thm:arbitrary_eavesdropped_lower_bound_noiseless_to_noisy} is not tight in general. As a result, we do not use it to bound all channels but instead apply it to a selective sequence of channels from $\mathcal{E}$. Notice that the model $\mathcal{C}_{\bar{e}}(R_c,0)$ for channel $\mathcal{C}_{\bar{e}}$ in Theorem~\ref{thm:arbitrary_eavesdropped_lower_bound_noiseless_to_noisy} sets the public rate $R_p$ to zero. This effectively removes $\bar{e}$ from all eavesdropping sets $E\in A$, giving a new adversarial set $A'=\big\{E\backslash\{\bar{e}\}:E\in A\big\}$. Repeated application of Theorem~\ref{thm:arbitrary_eavesdropped_lower_bound_noiseless_to_noisy} on a carefully chosen sequence of channels enable us to reduce all eavesdropping sets to size at most one. Once this is accomplished, we can use the equivalence result of Theorem~\ref{thm:lower_bound_one_or_no_eavesdropped} to replace the remaining noisy channels.

\begin{figure}
\centerline{\includegraphics[width=0.8\columnwidth]{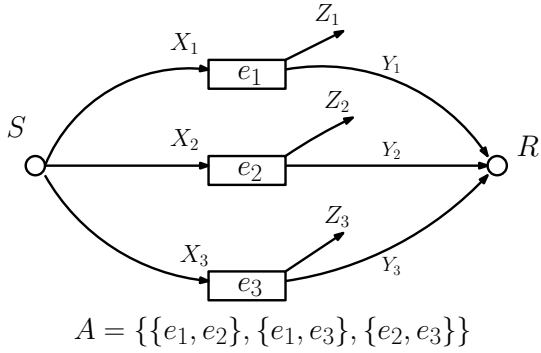}}
\caption{An example wiretap network for which lower bound model-II is not tight but lower bound model-II is tight.}
\label{fig:exp2}
\end{figure}

To show that lower bound model-I is not tight, consider the network of Figure~\ref{fig:exp2}, where each $i$ in $\{1,2,3\}$, $\max I(Y_i;X_i)=2$ and $\max I(Z_i;X_i)=1$. The adversary can  eavesdrop any two of $\{e_1,e_2,e_3\}$. Since for each link in  $\{e_1,e_2,e_3\}$ the confidential  capacity is $1$, and the public rate on two of the three links must be set to zero, the capacity of lower bound model-I is $3$. In the following we introduce lower bound model-II, using which we get a tighter lower bound, $4$, for this network.

\subsubsection*{Lower bound model-II}
\label{subsecsec:Model_noiseless_noisy_intuition}

\begin{figure}
\centerline{\includegraphics[width=0.6\columnwidth]{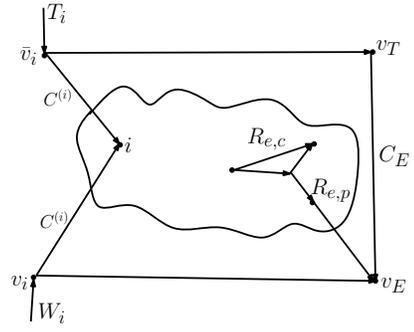}}
\caption{The $A$-enhanced network $\mathcal{N}(A)$.}
\label{fig:lower_bound_multiple_links_noiseless}
\end{figure}

In this model we bound the secrecy capacity region of network $\mathcal{N}$ with adversarial set $A\subseteq\mathcal{P}(\mathcal{E})$ by deriving a relationship 
with the traditional capacity of a noiseless communication network called the $A$-enhanced network $\mathcal{N}(A)$ defined below and illustrated by Figure~\ref{fig:lower_bound_multiple_links_noiseless}.
\begin{dfn}
Consider network $\mathcal{N}$ on graph $\mathcal{G}=(\mathcal{V},\mathcal{E})$. Define rate vector ${R}_{c,p}=\big((\check{R}_{e,c},{R}_{e,p}):e\in\mathcal{E}\big)$, and fix an adversarial set $A\subseteq\mathcal{P}(\mathcal{E})$. The $A$-enhanced network $\mathcal{N}({R}_{c,p},A)$ on graph $\check{\mathcal{G}}=(\check{\mathcal{V}},\check{\mathcal{E}})$ is defined as follows:
\begin{enumerate}
\item $\check{\mathcal{V}}=\mathcal{V}\cup\big\{v_i:i\in\mathcal{V}\big\}\cup\big\{\bar{v}_i:i\in\mathcal{V}\big\}\cup\big\{v_E:E\in A\big\}\cup\{v_T\}$. For each $i\in\mathcal{V}$ we call $v_i$ and $\bar{v}_i$ the $i^{\text{th}}$ message node and random key node of network $\mathcal{N}({R}_{c,p},A)$. For each $E\in A$, node $v_E$ is called an eavesdropper node. Node $v_T$ is called the overall key node.
\item $\check{\mathcal{E}}=\big\{h_i:i\in\mathcal{V}\big\}\cup\big\{\bar{h}_i:i\in\mathcal{V}\big\}\cup\big\{\check{\mathcal{C}}_e:e\in\mathcal{E}\big\}\cup\big\{h_e:e\in\mathcal{E}\big\}\cup\big\{(v_T,v_E,1):E\in A\big\}$.
\end{enumerate}

For each $i\in\mathcal{V}$, $h_i$ is a noiseless hyperarc of capacity ${C}^{(i)}$ (or alternatively a set of bit pipes each of capacity ${C}^{(i)}$) from node $v_i$ to all of the nodes in $\big\{i\big\}\cup\big\{v_E:E\in A\big\}$, and $\bar{h}_i$ is a noiseless hyperarc also of capacity ${C}^{(i)}$ (or alternatively a pair of bit pipes each of capacity ${C}^{(i)}$) from  node $\bar{v}_i$ to both of the nodes in $\big\{i, v_T\big\}$, where ${C}^{(i)}$ is defined in~(\ref{eq:outgoint_channel_capacity}) as the sum of the outgoing channel capacities from node $i$.

For each $e=(i,j,k)\in\mathcal{E}$, channel $\check{\mathcal{C}}_e$ in network is a bit pipe of capacity $R_{e,c}$ from node $i$ to node $j$, and hyperarc $h_e$ is a noiseless hyperarc of capacity $R_{e,p}$ from node $i$ to all of the nodes in $\big\{j\big\}\cup\big\{v_E: E\in A, e\in E\big\}$. For every $E\in A$ channel $\mathcal{C}_{(v_T,v_E,1)}$ is noiseless bit pipe of capacity
\begin{align*}
C_E=\sum_{i\in\mathcal{V}}C^{(i)}-\sum_{e\in E}{R}_{e,p}
\end{align*}
from node $v_T$ to node $v_E$.
\label{dfn:noiseless_network_lower_bound}
\end{dfn}

The $A$-enhanced network is used for traditional (rather than secure) communication with a collection of reconstruction constraints that depend on both $\mathcal{N}$ and $A$.
\begin{dfn}
Let $\mathcal{N}({R}_{c,p},A)$ be the $A$-enhanced network for network $\mathcal{N}$ and adversarial set $A\subseteq\mathcal{P}(\mathcal{E})$. A blocklength-$n$ solution $\mathcal{S}(\mathcal{N}({R}_{c,p},A))$ to network $\mathcal{N}({R}_{c,p},A)$ is defined as a set of encoding functions for each node $v$ in $\check {\mathcal V}$
\begin{align*}
(X^{(v)})^n &:(\mathcal{Y}^{(v)})_1^{n-1}\times (\mathcal{W}^{(v)})_1^{n-1} \times (\mathcal{T}^{(v)})_1^{n-1}  \longrightarrow (\mathcal{X}^{(v)})^n\end{align*}
and decoding functions
\begin{align*}
(\hat{W^{(v)}})^n &:(\mathcal{Y}^{(v)})_1^{n-1}\times (\mathcal{W}^{(v)})_1^{n-1} \times (\mathcal{T}^{(v)})_1^{n-1}\longrightarrow (\mathcal{W}^{(v)})\\
(\hat{T^{(v)}})^n &:(\mathcal{Y}^{(v)})_1^{n-1}\times (\mathcal{W}^{(v)})_1^{n-1} \times (\mathcal{T}^{(v)})_1^{n-1}\longrightarrow (\mathcal{T}^{(v)}).
\end{align*}
such that for each $i\in\mathcal{V}$ and $\mathcal{B}\in\mathcal{B}^{(i)}$, message  $W^{(v_i\rightarrow\mathcal{B})}$ from node $v_i$ is delivered to all of the nodes in $\mathcal{B}\in\mathcal{B}^{(i)}$, where $\mathcal{B}^{(i)}$ is the receivers set for node $i\in\mathcal{V}$ in network $\mathcal{N}$, and  random keys $T^{(i)}\in\mathcal{T}^{(i)}=\{1,\ldots,2^{n{C}^{(i)}}\}$ are delivered from node $\bar{v}_i$ to nodes $\{v_E:E\in A\}$.
\label{dfn:defining_a_solution_for_A_enhanced_network}
\end{dfn}
\begin{dfn}
The rate region $\mathcal{R}(\mathcal{N}({R}_{c,p}, A))\subseteq \mathbb{R}\hspace{0.4mm}^{m(2^{m-1}-1)}_+$ of the $A$-enhanced network $\mathcal{N}({R}_{c,p}, A)$ of network $\mathcal{N}$ is the closure of all rate vectors $R$ such that for any $\lambda>0$, a solution $(\lambda,R)$--$\mathcal{S}(\mathcal{N}({R}_{c,p},A))$ exists.
\label{dfn:defining_a_rate_region_for_A_enhanced_network}
\end{dfn}
\begin{theorem}
Consider network $\mathcal{N}$ on graph $\mathcal{G}=(\mathcal{V},\mathcal{E})$ and an adversarial set $A\subseteq\mathcal{P}(\mathcal{E})$. Let $\mathcal{N}({R}_{c,p}, A)$ be the $A$-enhanced network of network $\mathcal{N}$. If for every $e\in\mathcal{E}$
\begin{align*}
{R}_{e,p} &< \displaystyle\mathop{\max}_{p(x)} I(X^{(e)};Z^{(e)})\\
{R}_{c,p} &< \displaystyle\mathop{\max}_{p(x)} I(X^{(e)};Y^{(e)})- \displaystyle\mathop{\max}_{p(x)} I(X^{(e)};Z^{(e)}),
\end{align*}
then $\mathcal{R}(\mathcal{N}({R}_{c,p},A))\subseteq\mathcal{R}(\mathcal{N},A)$.
\label{thm:arbitrary_eavesdropped_lower_bound_noiseless_to_noisy_model_2}
\end{theorem}
\noindent{\it Sketch of the proof:}
We  start with a code for network $\mathcal{N}({R}_{c,p},A)$ and we will construct a secure code for network $\mathcal{N}$. 
We make use of an auxiliary network \textbf{I} which is the same as the A-enhanced network except that the noiseless bit pipes in $\big\{\check{\mathcal{C}}_e:e\in\mathcal{E}\big\}\cup\big\{h_e:e\in\mathcal{E}\big\}$ are changed back to the original noisy channels.  
We  show that we can emulate the   given code on network \textbf{I} such that the auxiliary receivers are still able to decode the required messages.  Since the total capacity of all incoming links to the auxiliary receivers is almost equal to the entropy of $(P^n, C^n, W,(Z^{E\backslash\{\bar{e}\}})^n)$, there is no spare capacity at links $((Z^{E\backslash\{\bar{e}\}})^n,Z^n)$ to carry any information about message $W$ and this corresponds to a secure code for network $\mathcal{N}$.
$\hfill\Box$

Unlike the rest of the results, where changing a single wiretap channel $\mathcal{C}_{\bar{e}}$ to its noiseless counterpart $\mathcal{C}_{\bar{e}}(R_c,R_p)$ results in an equivalent or bounding network, Theorem~\ref{thm:arbitrary_eavesdropped_lower_bound_noiseless_to_noisy_model_2} requires all wiretap channels in the noisy network $\mathcal{N}$ to be changed to noiseless channels in order to obtain a lower bounding network. Intuitively, this is because our construction requires the eavesdropper $E\in A$ to decode all sources of randomness in the network, which is not possible generally for noisy networks where the entropy of the noise can be potentially infinite. If we wish to replace only some noisy channels by their noiseless counterparts then   Theorem~\ref{thm:arbitrary_eavesdropped_lower_bound_noiseless_to_noisy} should be used. When all channels are to be replaced Theorem~\ref{thm:arbitrary_eavesdropped_lower_bound_noiseless_to_noisy_model_2} can be used, potentially leading to a tighter bound.

\begin{figure}
\centerline{\includegraphics[width=0.8\columnwidth]{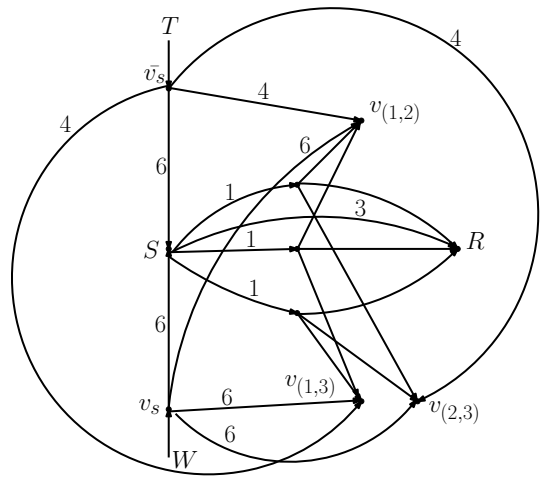}}
\caption{The A-enhanced network for the network in Figure~\ref{fig:exp2}. The number on top of each link represents the link capacity.}
\label{fig:exp_model_2}
\end{figure}

For example, we consider the network  in Figure~\ref{fig:exp2} where  model-I gives a lower bound of $3$. Here, we show that lower bound model-II gives a tighter lower bound, $4$. The A-enhanced network is shown in Figure~\ref{fig:exp_model_2}. For  simplicity, we combine the three direct links (with capacity 1) from $S$ to $R$ into a single link with capacity $3$.  The following code achieves rate  $(R_W,R_T)=(4,6)$  in the A-enhanced network. 
Let  $W=\{W_1,\ldots,W_4\}$ and $T=\{T_1,\ldots,T_6\}$.
The outgoing link of $S$ with capacity $3$ directly delivers $\{W_1,W_2,W_3\}$ to $R$. Each of other outgoing links of $S$ transmits a linearly independent combination of $\{W_4,T_5,T_6\}$. Node $\bar {V_S}$ transmits $\{T_1,T_2,T_3,T_4\}$ to each of $\{V_{\{1,2\}},V_{\{1,3\}},V_{\{2,3\}}\}$. Node $\{V_S\}$ transmits $\{W_1,\ldots,W_4\}$ to each of $\{V_{\{1,2\}},V_{\{1,3\}},V_{\{2,3\}}\}$. $R$ can decode $W_4$  from  the three linearly independent combinations of $\{W_4,T_5,T_6\}$. 
At $V_{\{1,2\}}$, messages $\{T_1,T_2,T_3,T_4\}$ and $\{ W_1,W_2,W_3,W_4\}$ are directly received from $\bar {V_S}$ and $V_S$, respectively. By using $W_4$  and  two linearly independent combinations of $\{W_4, T_5,T_6\}$, node $V_{\{1,2\}}$ can decode $\{T_5,T_6\}$. $V_{\{1,3\}},V_{\{2,3\}}$ decode similarly.

\section*{Acknowledgments}This work has been supported in part by NSF
grants CNS 0905615, CCF 0830666, and
    CCF 1017632.

\bibliographystyle{IEEEtran}
\bibliography{IEEEabrv,NWC-abbr-Ted-equ}
\end{document}